\documentclass[usegraphicx,usenatbib]{mn2e}

\newcommand\cen{Cen~X-4}

\newcommand\xmm{{\it XMM-Newton}}
\newcommand\swi{{\it Swift}}
\newcommand\cxo{{\it Chandra}}

\newcommand\ion[2]{#1$\;${\scshape{#2}}}

\begin{document}

\title[X-ray/UV in \cen]{An X-ray-UV correlation in Cen~X-4 during quiescence}

\author[Cackett et~al.]{E.~M.~Cackett$^1$\thanks{ecackett@wayne.edu},
E.~F.~Brown$^2$,
N.~Degenaar$^3$,
J.~M.~Miller$^3$,
M.~Reynolds$^3$,
\newauthor
R.~Wijnands$^4$
\\$^1$ Department of Physics \& Astronomy, Wayne State University, 666 W. Hancock St., Detroit, MI 48201, USA
\\$^2$ Department of Physics \& Astronomy, National Superconducting Cyclotron Laboratory, and the Joint Institute for Nuclear Astrophysics,\\ Michigan State University, East Lansing, MI 48824, USA
\\$^3$ Department of Astronomy, University of Michigan, 500 Church St, Ann Arbor, MI 48109-1042, USA
\\$^4$ Astronomical Institute `Anton Pannekoek', University of Amsterdam, Science Park 904, 1098 XH Amsterdam, the Netherlands
}
\date{Received ; in original form }
\maketitle

\begin{abstract}

Quiescent emission from the neutron star low-mass X-ray binary Cen~X-4 is seen to be variable on timescales from hundreds of seconds to years, suggesting that at least in this object, low-level accretion is important during quiescence.  Here we present results from recent \xmm\ and \swi\ observations of Cen~X-4, where the X-ray flux (0.5 -- 10 keV) varies by a factor of 6.5 between the brightest and faintest states.  We find a positive correlation between the X-ray flux and the simultaneous near-UV flux, where as there is no significant correlation between the X-ray and simultaneous optical (V, B) fluxes.  This suggests that while the X-ray and UV emitting regions are somehow linked, the optical region originates elsewhere.  Comparing the luminosities, it is plausible that the UV emission originates due to reprocessing of the X-ray flux by the accretion disk, with the hot inner region of the disk being a possible location for the UV emitting region.  The optical emission, however, could be dominated by the donor star.  The X-ray/UV correlation does not favour the accretion stream-impact point as the source of the UV emission.
\end{abstract}


\section{Introduction}

Low-mass X-ray binaries are often transient, cycling through periods of outburst and quiescence where the X-ray luminosity varies by over four orders of magnitude between the two.  In quiescence, the accretion rate is at less than $10^{-4}$ of the Eddington limit.  The transient behavior of X-ray binaries (and dwarf novae) is broadly described by the disk instability model \citep[DIM, see][for a review]{lasota01, lasota08} where a thermal-viscous instability causes an outburst.  During an outburst accretion likely occurs via an optically-thick, geometrically-thin disk  \citep[e.g.][]{shakurasunyaev73}.  During quiescence, the picture is not quite as clear, and accretion is thought to occur in a very different manner.  In order to describe both the long recurrence times for transients as well as the observed quiescent X-ray emission a standard thin disk must be truncated at large radii \citep[e.g.][]{lasota96, dubus01}.  Moreover, as the accretion rate drops, it is expected that the inner disk evaporates, leaving a hole \citep{meyer94}.

Two components models for the quiescent accretion flow were therefore developed, where an outer truncated disk has an advection-dominated accretion flow \citep[ADAF;][]{narayanyi94,narayanyi95} inside \citep{narayan96,narayan97, esin97}.  Such models successfully explain the quiescent X-ray properties of black hole LMXBs, which are seen to display simple power-law emission.  For neutron stars, there is the added complication of a stellar surface.  The ADAF models  predicts that neutron stars should have a higher quiescent luminosity compared to black holes (with the same orbital period) as the hot flow can heat the neutron star surface, where as in black holes the energy will be advected through the event horizon \citep[e.g.,][and references therein]{garcia01}.  While such a difference is generally seen \citep[though see][for a remarkable exception]{jonker07} it is also expected that quiescent neutron stars should be hot due to heating of the crust during the outburst \citep[`deep crustal heating';][]{BBR98}.  Moreover, it is even possible that energy may go into a jet rather than being advected \citep{fender03}.

The exact geometry and details of how the hot, radiatively inefficient accretion flows work during quiescence is still not clear.  There is uncertainty as to whether the flow is an ADAF \citep{narayanyi94,narayanyi95}, or an advection-dominated flow with strong outflows \citep[ADIOS][]{blandbegel99}, and the value of the standard viscosity parameter, $\alpha$,  in quiescence \citep{menou00,menou02}. The truncation radius is usually assumed to be between $10^2$ and $10^4$ Schwarzschild radii \citep{esin97}, but it too is uncertain  \citep{menou99}.  How matter is accreted onto the neutron star is also not clear.  The quiescent luminosities of neutron star LMXBs are low enough to require that much of the accreted matter is prevented from reaching the neutron star surface, which could happen via the propeller effect \citep{menou99,menoumcclintock01}.  Alternatively, the hot accretion flow may not be an ADAF (or ADIOS), but a hot settling flow instead \citep{medvedev01,narayan03}.  In this hot settling flow scenario, hot quasi-spherical accretion onto a spinning neutron star happens subsonically, and the flow essentially settles onto the rotating neutron star without a shock.  The luminosity of this flow is then mostly generated as rotational energy is extracted from the neutron star, rather than from the mass accretion rate.  Thus, there exists a range of models to explain how accretion happens during quiescence, and even the basic parameters of such models are not well constrained, at least in the case of neutron star LMXBs.

The mechanism behind accretion in quiescence therefore remains elusive.  Yet, one signature that it must be occurring is sporadic X-ray variability.  Such variability in 
quiescent emission from neutron stars has been seen in several objects on all timescales studied -- from hundreds of seconds to years \citep{campana97,campanaetal04, rutledgeetal01a, rutledge02, cackett05, cackett10,cackett11,fridriksson11}.  X-ray spectra of quiescent neutron stars typically show two components - both a thermal, blackbody-like component and a non-thermal power-law component.  A recent study of long-term quiescent variability in  Cen~X-4 \citep{cackett10} confirmed that the thermal component, as well as the power-law component, is variable.  Interestingly, the power-law and thermal fluxes appear to vary in tandem, contributing roughly the same fraction of the total flux at all epochs.  This suggests a clear link between the two components during quiescence.  Compton upscattering of soft photons from the neutron star surface in the hot accretion flow would seem to be an obvious link between these tow components.  However, it would produce a power-law that is only a few percent of the 0.5 -- 10 keV flux \citep{menoumcclintock01}, while we observe the components to be roughly equal \citep{cackett10}.  The power-law arising from shock emission between a radio pulsar wind and inflowing matter from the companion star \citep{campanastella03} also seems to be ruled out as the variability in \cen\ cannot be explained by correlated changes in column density and the power-law component \citep{cackett10}. The accretion flow getting all the way down to the neutron star surface may therefore be the most realistic mechanism for the synchronous change in both the neutron star surface temperature and the power-law component as the accretion rate varies \citep[e.g.][]{stellaetal94, menoumcclintock01}. 

In order to further investigate the nature of quiescent variability in neutron star low-mass X-ray binaries on timescales of weeks to months we observed \cen\ with \xmm\ four times between August 2010 and January 2011.  Importantly, we also obtain optical and UV photometry with \xmm\ to investigate the nature of broadband variability during quiescence.  Combining these four new \xmm\ observations with one archival \xmm\ observation and two \swi\ observations we show a clear correlation between X-ray and UV flux during quiescence.  \cen\ is one of very few objects where such an X-ray/UV study can be performed given its proximity and relatively low extinction.  In Section ~\ref{sec:data} we describe the data reduction and analysis before describing our results in Section~\ref{sec:result} and discussing their implications in Section~\ref{sec:discuss}.

\section{Data Reduction and Analysis} \label{sec:data}

The four \xmm\ observations were obtained between the end of August 2010 and January 2011 with each observation lasting between 10 and 20 ksec.  The ObsIDs are 0654470201, 0654470301, 0654470401, 0654470501 and their respective start dates (dd/mm/yy) are 25/08/10, 04/09/10, 24/01/11 and 31/01/11.   The X-ray detectors were operated in full window mode with a medium filter.  The Optical Monitor \citep[OM,][]{mason01} was operated such that data was collected in multiple optical/UV filters during the X-ray observations.  In Table~\ref{tab:obs} we give details of all four observations along with exposure times for each instrument and filter used.  The data was reduced with the \xmm\ Science Analysis Software, version 11.0.0, using the latest calibration files.   

In addition to the \xmm\ data, we also analyze two \swi\ observations of \cen\ that also have both X-ray and optical/UV data.  The ObsIDs 00035324001 and 00035324002 were performed on  03/09/06 and 01/05/12.  The X-ray detector was operated in photon counting mode.  In the first observation, only a UVW1 exposure was taken, where as in the second observation exposures in V, B, U, UVW1, UVM2, and UVW2 were taken.  Table~\ref{tab:swi} gives details of the \swi\ observations.  The data was reduced using HEASOFT, version 6.11.1.  In the following sections we detail the data reduction for the X-ray and optical/UV detectors.

\begin{table*}
\centering
\caption{Details of the new \xmm\ observations of Cen~X-4, including count rates and optical/UV fluxes. AB flux densities have units of erg s$^{-1}$ cm$^{-2}$ \AA.  ND indicates that the source was not detected in that filter. MOS and PN net count rates are given in the 0.5 -- 10 keV energy range.  The count rate is given for MOS 1 only, the MOS 2 rate is always very similar.}
\label{tab:obs}
{\scriptsize
\begin{tabular}{lcccccccccccc}
\hline
Detector  & \multicolumn{3}{c}{0654470201}  & \multicolumn{3}{c}{0654470301} & \multicolumn{3}{c}{0654470401} & \multicolumn{3}{c}{0654470501} \\
or  & Exp. & Net rate & AB flux &  Exp. & Net rate & AB flux &  Exp. & Net rate & AB flux &  Exp. & Net rate & AB flux\\
filter & (ks) & ($10^{-1}$  c/s) & $(10^{-17})$ & (ks) & ($10^{-1}$  c/s) & $(10^{-17})$ & (ks) & ($10^{-1}$  c/s) & $(10^{-17})$ & (ks) & ($10^{-1}$ c/s) & $(10^{-17})$\\
\hline
MOS & 11.2 & $0.75\pm0.03$ & -- & 19.3 & $0.84\pm0.02$ &--  & 14.4 & $1.20\pm0.03$ & --& 13.4 & $0.39\pm0.02$ & -- \\ 
PN & 10.4 & $3.07\pm0.07$ & -- & 16.1 &  $3.73\pm0.05$ &-- & 11.3 &$4.21\pm0.06$ & --& 10.5 & $1.59\pm0.04$ & --  \\ 
V & 4.0 & $6.2\pm0.6$ & $15.5\pm1.4$ & 2.86 & $9.2\pm0.7$ & $23.1\pm1.8$ & 1.76 & $5.1\pm0.8$ & $12.8\pm2.1$ & 1.76 &  $4.8\pm0.8$ & $12.0\pm2.0$  \\
B & 3.0 & $7.0\pm0.8$ & $8.7\pm1.0$ & 0.0 & --  & -- & 1.76 & $8.2\pm1.0$ & $10.2\pm1.2$ & 1.76 & $6.4\pm1.0$ & $8.0\pm1.2$ \\
U & 1.98 & $2.9\pm0.6$ & $5.6\pm1.2$ & 3.76 & $4.6\pm0.5$ & $8.8\pm0.9$ &1.76 & $5.0\pm0.6$ & $9.7\pm1.2$ & 1.76 & $2.6\pm0.6$ & $5.0\pm1.2$\\
UVW1 & 5.0 & $1.7\pm0.2$ & $8.3\pm1.1$ & 3.76 & $1.8\pm0.3$ & $8.8\pm1.3$ & 1.76 & $2.4\pm0.4$ & $11.5\pm2.0$ &1.76 & $0.42\pm0.36$ & $2.0\pm1.7$ \\
UVM2 & 5.0 & $0.37\pm0.14$ & $8.2\pm3.0$ & 5.0 & ND & ND & 4.2 & $0.57\pm0.15$ & $12.6\pm3.3$ & 3.2 & ND & ND \\
\hline
\end{tabular}}
\end{table*}

\begin{table*}
\centering
\caption{Details of the \swi\ observations of Cen~X-4, including count rates and optical/UV fluxes. AB flux densities have units of erg s$^{-1}$ cm$^{-2}$ \AA.  XRT net count rates are given in the 0.5 -- 10 keV energy range.}
\label{tab:swi}
\begin{tabular}{lcccccc}
\hline
Detector  & \multicolumn{3}{c}{00035324001}  & \multicolumn{3}{c}{00035324002}  \\
or  & Exp. & Net rate & AB flux &  Exp. & Net rate & AB flux \\
filter & (ks) & ($10^{-1}$  c/s) & $(10^{-17})$ & (ks) & ($10^{-1}$  c/s) & $(10^{-17})$ \\
\hline
XRT & 4.5 & $0.38\pm0.03$ & -- & 3.8 & $0.76\pm0.05$ & --  \\
V  & -- & -- & -- & 0.30 & $0.75\pm0.07$ & $19.5\pm2.0$  \\
B  & -- & -- & -- & 0.30 & $1.11\pm0.10$ & $16.5\pm1.4$ \\
U  & -- & -- & -- & 0.30 & $1.35\pm0.09$ & $22.5\pm1.5$ \\
UVW1 & 6.25 & $0.32\pm0.01$ & $15.0\pm0.5$ & 0.61 & $0.55\pm0.04$ & $25.6\pm1.9$ \\
UVM2 & -- & -- & -- & 0.95 & $0.29\pm0.02$ & $24.4\pm1.8$\\
\hline
\end{tabular}
\end{table*}

\subsection{X-ray data reduction}

\subsubsection{\xmm}
We created calibrated event files for each observation from the Observation Data Files using the emproc and epproc commands for the MOS and PN detectors.  We check for periods of high background by creating light curves from the entire detectors with 100s time binning, filtering for events with energies $>10$ keV and PATTERN=0 for the MOS, and $10 < E < 12$ keV and PATTERN=0 for the PN.

The background was high throughout the entire first observation (0654470201), with the MOS1 count rate ranging from 0.2 -- 3.7 c/s, and the PN count rate from 2.6 -- 15.2 c/s.  Filtering out the periods of particularly high background would leave only a very small exposure left, thus we include all the available data.  The other three observations had some short periods of background flaring at either the beginning or end  of the observations, but the background was mostly low throughout.  The periods of background flaring were never as bad as the first observation, and therefore we include all available data.  Only in the first observation does the significant flaring there noticeably reduce the S/N ratio.

For both the MOS and PN detectors, we extract the source spectrum from a circular region of radius 20 arcsec, and the background spectrum from a nearby, source-free 2 arcmin region.  The response files are generated with the arfgen and rmfgen tools, and the resulting spectrum is binned to a minimum of 25 counts per bin.  We give net count rates in Table~\ref{tab:obs}.

\subsubsection{\swi}

Calibrated event files were created by reprocessing the data using the xrtpipeline tool, and applying the standard (default) screening criteria.  We analyze data taken in photon counting mode.  We extracted the source spectrum using xselect, and a circular extraction region with a 20 pixel radius.  The background spectrum was extracted from an annulus with inner radius of 40 pixels and outer radius of 120 pixels.  We use the xrtmkarf tool to create the ancillary response file, and use the appropriate response matrix based on the epoch of the observation and the observing mode.  Given the shorter exposures and lower count rates compared to \xmm\ we only bin spectra to 10 counts per bin.

\subsection{Optical/UV data}

\subsubsection{\xmm}
We observed Cen~X-4 with the OM in imaging mode.  However, rather than use the standard set-up, we chose to more efficiently obtain exposures of Cen~X-4 in 5 filters by using a Science User Defined mode whereby only a single window is used to observe just the center of the field of view (normally a mosaic is created from multiple windows to cover a much larger field of view).  In this way, we were able to obtain images in the following filters during the short X-ray exposures of Cen X-4 (effective wavelengths are given): V (5407\AA), B (4334\AA), U (3472\AA), UVW1 (2905\AA) and UVM2 (2298\AA).  

We reduce the OM data using the omichain tool.  This tool processes the data with the latest calibration files, and then performs source detection and aperture photometry.  The output files include both the images for each filter along with a combined source list containing detailed parameters for each source detected in all exposures, including the count rate (corrected for instrumental effects such as detector deadtime), and AB flux densities (in erg s$^{-1}$ m$^{-2}$ \AA$^{-1}$).  We visually identified Cen X-4 in all images before finding the corresponding source in the source list.  In observation 0654470301 Cen X-4 was not automatically detected in the UVM2 filter, though a faint excess in counts did seem apparent on visual inspection.  We therefore ran the omdetect tool (part of the omichain pipeline) on its own with the significance reduced to only requiring 1$\sigma$ above the background, however, the source was still not detected (note the default is a 3$\sigma$ threshold).  Similarly, in observation 0654470501 the source was not automatically detected in UVW1 or UVM2.  Once again, it did seem present (though faint) on manual inspection of the image, and we therefore re-ran the omdetect tool with only a 1$\sigma$ threshold.  This led to a detection of Cen X-4 in UVW1 with only a 1.9$\sigma$ significance, while there was still no positive detection in the UVM2 filter.  Count rates and AB flux densities are given for each observation in Table~\ref{tab:obs}.  There is clearly variability seen between the different observations.

We also note that only one of the two archival \xmm\ observations (ObsID:0144900101) also has OM data.  This observation took place starting on 1 March 2003, with exposures in V, U, B, UVW1, UVM2 and UVW2 filters.  The data was taken in imaging mode, using the default setup whereby a series of 5 exposures are taken in each filter, and then combined together in a  mosaic to fill the 17 arcmin field of view.  During each exposure the central 2 arcmin region is always observed.  We reduced the data in the same manner, using the omichain tool once again.  The exposure times,  corrected count rates and fluxes for each filter are given in Table~\ref{tab:oldxmm}.

\begin{table}
\centering
\caption{\xmm\ obsID 0144900101, optical/UV details. AB flux densities have units of erg s$^{-1}$ cm$^{-2}$ \AA.}
\label{tab:oldxmm}
\begin{tabular}{lccc}
\hline
Filter  &  Exp. & Net rate & AB flux \\
 & (ks) & ($10^{-1}$  c/s) & $(10^{-17})$ \\
\hline
V & 4.0 & $8.3\pm0.5$ & $20.8\pm1.2$ \\
B & 4.0 & $13.9\pm0.5$ & $17.3\pm0.7$\\
U & 4.0 & $6.6\pm0.3$ & $12.7\pm0.7$  \\
UVW1 & 12.2 & $3.7\pm0.1$ & $18.0\pm0.5$ \\
UVM2 & 19.8 & $0.81\pm0.05$ & $17.9\pm1.0$ \\ 
UVW2 & 25.0 & $0.25\pm0.07$ & $14.5\pm3.8$\\
\hline
\end{tabular}
\end{table}

\subsubsection{\swi}

The two \swi\ observations also include optical/UV observations using the UVOT \citep{roming05,breeveld10}.  The filters are similar, though slightly different, to the \xmm\ filters. The first observation used the UVW1 filter exclusively, where as the second observation used V (5468\AA), B (4392\AA), U (3465\AA), UVW1 (2600\AA), UVM2 (2246\AA) and UVW2(1928\AA) filters \citep[wavelengths from][]{poole08}.  Note that both the UVW1 and UVW2 filters are broader than the UVM2 filter, and hence are more sensitive. However, UVW1 and UVW2 also have red leaks, meaning that their sensitivity stretches to longer (redder) wavelengths than UVM2.  We used the uvotmaghist tool to perform aperture photometry, using a 3 arcsec radius circular extraction region, and a circular, source-free nearby background region with a 12 arcsec radius.  This tool performs photometry on all separate exposures for a given filter, applies the aperture correction and determines the AB flux.  Having determined the count rates and fluxes for each individual exposure, we calculate the exposure-weighted average count rate and flux, as given in Table~\ref{tab:swi}.

\subsection{X-ray Spectral analysis}

We fit the X-ray spectra using XSPEC version 12.7.0 \citep{arnaud96}.  Each observation is fit separately, but when fitting the individual \xmm\ observations we jointly fit the MOS 1, MOS 2, and PN spectra, with the parameters the same for all detectors.  We fit the spectra using a neutron star atmosphere plus a power-law, all modified by Galactic photoelectric absorption.  We use the nsatmos model for the neutron star atmosphere \citep{heinke06}.  This is the same model we used to fit six previous observations of \cen\ \citep{cackett10}, thus allowing for a direct comparison with these previous results\footnote{This previous analysis looked at data from a range of different missions (1 {\it ASCA}, 2 {\it Chandra}, 2 {\it XMM} and 1 {\it Suzaku})}.  For the same reason, we choose to fix $N_{\rm H} = 4.9\times10^{20}$ cm$^{-2}$, the value we found from jointly fitting six previous observations in \citet{cackett10}.  Also, to allow a direct comparison with our previous results, we fix the distance to 1.2 kpc \citep{chevalier89} in the spectral fits.  In \citet{cackett10} we investigated which parameters were variable, finding that the thermal component and the power-law component must both vary.  For the thermal component, either the effective temperature, emitting radius or both can vary, and we found that equally good fits are achieved regardless of which parameter is left variable.  Here, for simplicity, we fix the neutron star radius and allow the temperature to vary between epochs.  This choice does not affect the results since the thermal component can be fit equally well with either a variable temperature or a variable radius, and it is only the flux that we are mostly concerned with here.  We assume a neutron star with $R = 10$ km and $M = 1.4$ M$_\odot$, and that the entire surface is emitting.  We fit the spectra in the 0.5 -- 10 keV range and best fitting parameters are given in Table~\ref{tab:specfit}.  Uncertainties are quoted at the 1$\sigma$ level throughout.

\begin{table*}
\centering
\caption{Spectral fit parameters.  The flux is given in units of erg s$^{-1}$ cm$^{-2}$. The column density was fixed at $N_{\rm H} = 4.9\times10^{20}$ cm$^{-2}$ in all spectral fits.  We also assumed a neutron star radius of 10 km, and mass of 1.4 M$_\odot$, and a distance to Cen~X-4 of 1.2 kpc.  The power-law normalization is defined as photons keV$^{-1}$ cm$^{-2}$ s$^{-1}$ at 1 keV.  We define the thermal fraction is the ratio of the unabsorbed 0.5 -- 10 keV thermal flux to the total unabsorbed 0.5 -- 10 keV flux.}  
\label{tab:specfit}
\begin{tabular}{lcccccc}
\hline
 & \multicolumn{4}{c}{\it XMM-Newton} & \multicolumn{2}{c}{\it Swift} \\
Parameter  & 0654470201 & 0654470301 & 0654470401 & 0654470501 & 00035324001 & 00035324002 \\
\hline
$kT_{\rm eff}^{\infty}$ (eV)     & $55.9\pm0.7$ & $56.5\pm0.3$ & $60.8\pm0.3$ & $49.0\pm0.6$ & $62.4\pm2.3$ & $65.2\pm5.7$ \\
Power-law index, $\Gamma$ & $1.77\pm0.21$ & $1.62\pm0.10$ &  $1.38\pm0.10$ & $1.94\pm0.19$ & $1.51\pm0.86$ & $1.79\pm0.53$\\
Power-law norm ($10^{-5}$) & $6.6\pm1.4$ & $6.3\pm0.7$ & $5.5\pm0.7$ & $3.6\pm0.7$ & $5.8^{+7.5}_{-4.8}$ & $29.1^{+22.4}_{-13.5}$ \\
Unabs. 0.5 -- 10 keV flux ($10^{-12}$) & $0.92\pm0.08$ & $1.00\pm0.02$ & $1.31\pm0.03$ & $0.44\pm0.01$ & $1.38\pm0.38$ & $2.85\pm0.30$ \\
Thermal fraction (\%) & 58 & 56 & 62 & 59 & 68 & 41  \\
$\chi^2_\nu$ (dof) & 0.91 (106) & 0.93 (168) & 0.99 (129) & 1.27 (71) & 1.31 (13)  & 0.67 (24)\\
\hline
\end{tabular}
\end{table*}

\section{Results} \label{sec:result}

\begin{figure}
\centering
\includegraphics[angle=270,width=8.4cm]{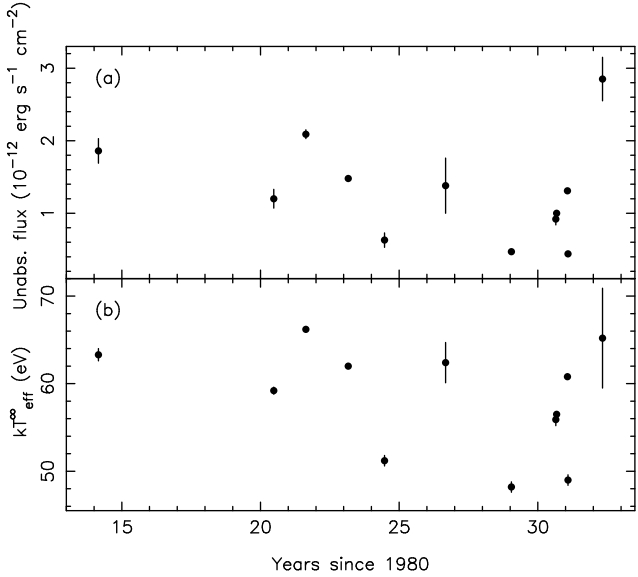}
\caption{Long-term X-ray quiescent light curve of \cen, showing variability on all timescales. (a) Unabsorbed 0.5 -- 10 keV flux, (b) Effective temperature for an observer at infinity.}
\label{fig:lc}
\end{figure}

The long-term X-ray quiescent light curve of \cen\ is shown in Figure~\ref{fig:lc}, where we include the observations from \citet{cackett10} as well as the four new \xmm{} and two \swi{} observations presented here.  Notably, significant variability is seen on the weeks to months timescales probed by these new observations.  Most significant is the factor of 3 drop in flux between the third and forth observations taken at the end of January 2011 and separated by only 7 days.

\begin{figure}
\centering
\includegraphics[angle=270,width=8.4cm]{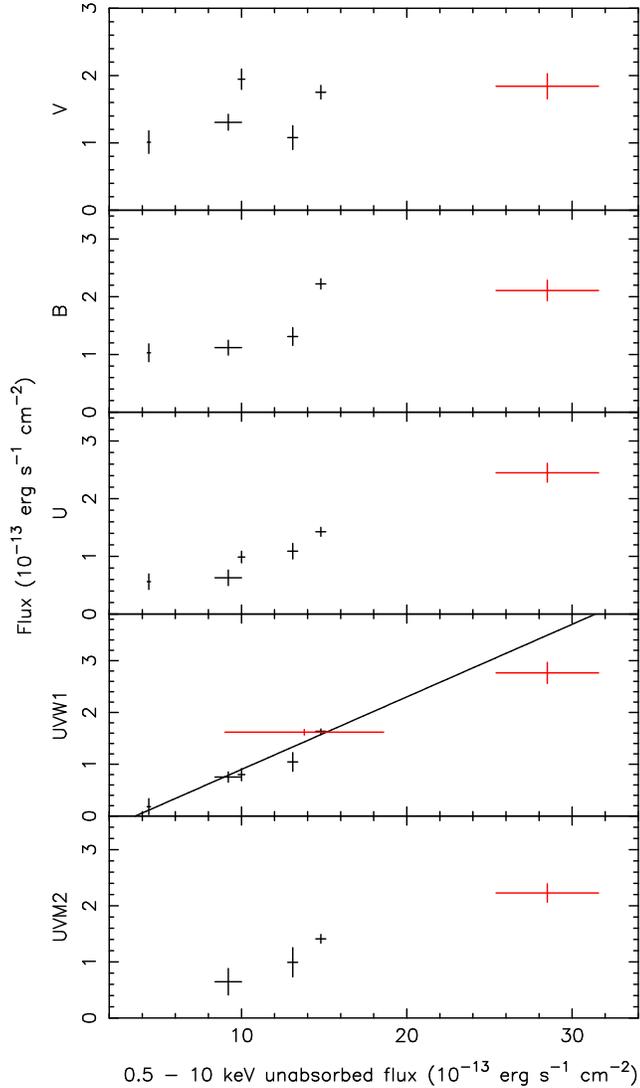}
\caption{De-reddened optical and UV fluxes versus 0.5 -- 10 keV unabsorbed flux from \xmm\ (black ) and \swi\ (red).  A clear correlation between the X-ray and the U, UVW1 and UVM2 fluxes is apparent.  The solid line shows the best-fitting straight line to the UVW1/X-ray points, with equation  $f_{UV} = (0.14\pm0.01) f_{X} - (0.50\pm0.18)$.}
\label{fig:optuv}
\end{figure}

It is clear from looking at the OM count rates and fluxes (Table~\ref{tab:obs}) that Cen~X-4 is also highly variable in the optical/UV part of the spectrum during quiescence.  In Figure~\ref{fig:optuv} we plot the optical/UV fluxes against the 0.5 -- 10 keV unabsorbed X-ray flux.  In order to compare the energetics, we convert the flux densities (erg s$^{-1}$ m$^{-2}$ \AA$^{-1}$) that are the output from the omichain to fluxes by multiplying by the FWHM of the filter bandpass.  We use:  684 \AA\ (V),  976\AA\ (B), 810\AA\ (U), 620\AA\ (UVW1), 439\AA\ (UVM2) for the \xmm\ filters, and 769\AA\ (V), 975\AA\ (B), 785\AA\ (U), 693\AA\ (UVW1), 498\AA\ (UVM2) for the \swi\ filters.  We also de-redden the UV/optical fluxes.  We use the gas-to-dust ratio from \citet{guverozel09} in order to convert from equivalent hydrogen column to $E(B-V)$.  We then use the interstellar extinction curve of \citet{cardelli89} to estimate the extinction correction at the wavelengths of each filter.

We test for a correlation between the optical/UV and X-ray fluxes using a simple Pearson's linear correlation test.  The correlation coefficients we find are $r = 0.554, 0.779, 0.982, 0.973, 0.976$ for the V, B, U, UVW1 and UVM2 filters respectively.  Given the number of data points, and a two-tail test (no a priori knowledge of positive or negative correlation), this corresponds to a positive correlation at the $0.746, 0.880, 0.99952, 0.99977,  0.976$ confidence levels.  Thus, the U and UVW1 fluxes are correlated with the X-ray flux at greater than 3$\sigma$, while the UVM2 flux is correlated at greater than 2$\sigma$.  There is no significant correlation between the V or B fluxes and the X-ray flux, indicating that this is a UV only correlation.  We also look to see if there is a correlation between UVW1 and V band fluxes, finding $r = 0.63$, which corresponds to a positive correlation at the 0.816 confidence level.  Thus, there is no significant correlation between the V band and UVW1 fluxes.

We now concentrate on the UVW1 versus X-ray correlation further, given that it is the most significant correlation.  Obviously, given the high linear correlation coefficient, a simple straight-line fits the UVW1 -- X-ray correlation well.  We get the following best-fitting parameters: $f_{UV} = (0.14\pm0.01) f_{X} - (0.50\pm0.18)$, giving a reduced-$\chi^2 = 1.35$.   We also test fitting a power-law of the form $f_{UV} = a(f_{X} - f_{X0})^b$, where $f_{X0}$ is a constant to allow a non-zero X-ray flux when $f_{UV} = 0.0$, or vice versa.  The best-fit is very close to a linear relationship, with $b = 1.00^{+0.34}_{-0.14}$, and giving a reduced-$\chi^2 = 1.69$.  If we fit a power-law with the index, b fixed at 0.5 \citep[as expected for reprocessing by][]{vanparadijs94} we get a worse fit (reduced $\chi^2 = 4.5$).
However, with only 7 data points we can clearly not make any strong conclusions about the form of the correlation, especially as there is only one observation at a high luminosity, where a deviation from a linear relationship would become apparent.  We also caution that we are comparing UVW1 fluxes from both {\it Swift} and {\it XMM-Newton} - the filter responses and wavelength range are slightly different between the missions which could lead to a small offset between the two.  Further data is needed to strengthen and define the shape of the correlation.

\begin{figure}
\centering
\includegraphics[angle=270,width=8.4cm]{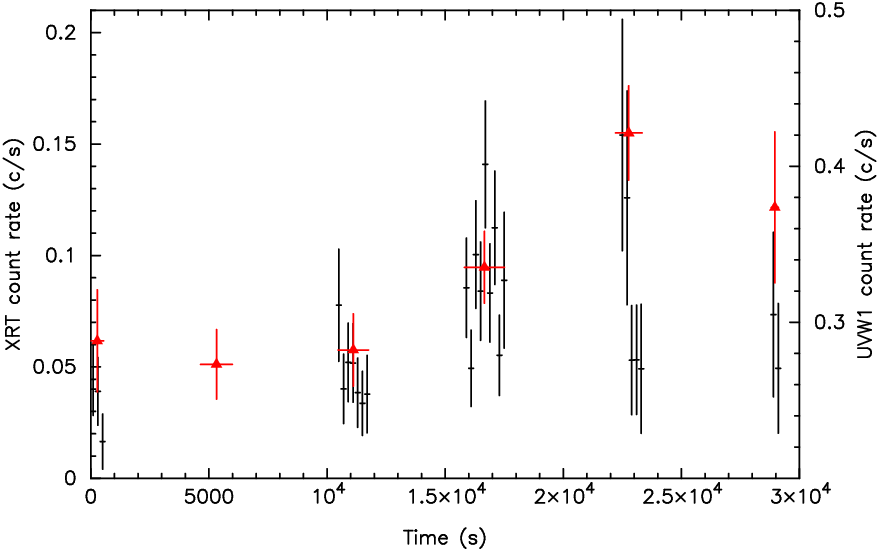}
\includegraphics[angle=270,width=8.4cm]{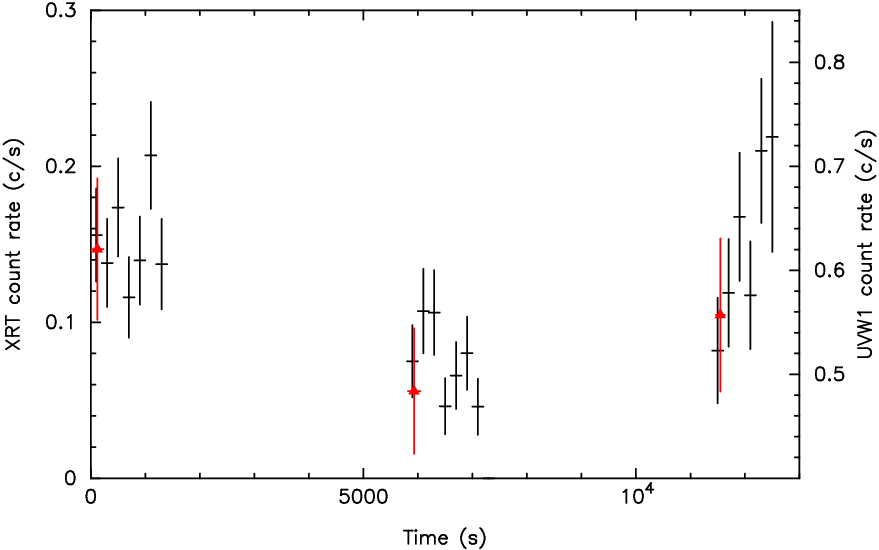}
\caption{\swi\ X-ray (black) and UVW1 (red, triangles) light curves during the first (top) and second (bottom) \swi\ observations (00035324001 and 00035324002). The XRT data has 200 second time binning.}
\label{fig:swiuvxray}
\end{figure}

We also look for whether there is an X-ray/UV correlation on shorter timescales by looking at the UV and X-ray light curve during the two \swi\ observations.  The is possible with \swi\ because the observations consist of multiple $\sim1$ ksec exposures, each giving separate UV images.  With \xmm\, however, there is only one exposures in each filter during the entire observation and so we cannot look for X-ray-UV correlations during the \xmm\ observations.  The first \swi\ observation consists of 6 \swi\ orbits of data, and the UVW1 filter was used for all of each orbit.  We can therefore compare the X-ray and UV light curves on timescales of a few thousand seconds.  The second \swi\ observation comprised three orbits, with a UVW1 exposure during each orbit.  For both observations, we create a background-subtracted \swi/XRT light curves, using the xrtlccorr tool to perform count rates corrections for vignetting, PSF etc.  The UVW1 count rates are determined using the uvotmaghist tool (as described above).  We show the light curves  in Figure~\ref{fig:swiuvxray}.  The XRT and UVW1 count rates are generally well correlated during both observations.

The \xmm\ observations only perform one exposure in each filter during a given observation.  Therefore, while the X-ray flux we obtain is from the entire observation, the optical/UV fluxes are only obtained during some fraction of the observation.  The X-ray exposures range from approximately 10 -- 20 ksec, where as the optical/UV exposures are between 1.76 and 5 ksec.  Variability during the observation could therefore add scatter to the UV and X-ray correlations.  Examining the X-ray lightcurves from the \xmm\ observations there is variability present \citep[as has also been seen during previous \xmm\ and \cxo\ observations, see, e.g.,][]{campanaetal04,cackett10}.  However, any subset of 2 ksec or longer will not deviate too far from the mean X-ray flux, and thus not add a large amount of scatter.

\section{Discussion and Conclusions} \label{sec:discuss}

Our observations of \cen\ during quiescence have shown that the X-ray and near-UV emission are correlated, where as there is no significant correlation between the X-ray and optical emission.  The X-ray/UV correlation is seen both on timescales of weeks -- years, as well as during observations spanning less than one day.  In order to investigate the physical origin of this correlation, it is important to consider the energetics -- if the correlation is due to reprocessing of the X-ray emission, then the near-UV flux must be only a fraction of the X-ray flux.

We find that the UVW1 flux is approximately 10\% of the 0.5 -- 10 keV flux.  However, much of the X-ray flux from the hot neutron star surface is emitted below 0.5 keV, and thus it is important to take this flux into consideration.  Furthermore, the power-law component will likely extend beyond 10 keV, though its shape there is uncertain.  To provide an estimate for the total X-ray flux, we extrapolate the best-fitting neutron star atmosphere plus power-law models to cover the range 0.01 -- 100 keV.  For the faintest observation (\xmm: 0654470501) we find a factor of 3 increase in the flux, where as for the brightest observation (\swi: 00035324002) we find a factor of 2.6 increase.  The higher temperature during the brighter observation means more of the neutron star atmosphere flux is in the 0.5 -- 10 keV band, hence the slightly lower increase in flux when extending the energy range.  Of course, the near-UV emission is in more than just the UVW1 filter.  Combining the U, UVW1 and UVM2 fluxes, we get an increase by a factor of 4.0 and 2.7 for the faintest and brightest observations respectively, over using just the UVW1 flux.  Thus, when accounting for emission over a wider wavelength range, the near-UV flux is still approximately 10\% of the X-ray flux.  Such a fraction of the X-ray flux is reasonable for reprocessed emission.

\citet{vanparadijs94} consider X-ray reprocessing in the accretion disk and show that the reprocessed emission should be proportional to $L_X^{1/2}R$, where $R$ is the outer radius of the accretion disk.  We, however, find that the UV flux increases approximately linearly with the X-ray flux and a power-law with index of 0.5 does give a significantly worse fit.  However, with the limited number of observations and range in flux the exact relationship is uncertain, and would be significantly strengthened by further data.  Considering the relevant timescales for reprocessing, the light-travel time from the neutron star to the outer disk is less than one second, and so we would not expect to see any time delay between the X-ray and UV lightcurves.  This is what we observe (Fig.~\ref{fig:swiuvxray}) where the X-ray and UV vary in tandem during {\it Swift} observations.  The viscous timescale from the outer disk in these objects is significantly longer (of the order of weeks), and therefore cannot be used to explain the simultaneous variability in UV and X-rays.

Another source where a similar correlation has been found is the quiescent black hole V404~Cyg, where a correlation between X-ray and H$\alpha$ emission has been previously observed \citep{hynes04}, with a weaker correlation between the X-ray flux and the optical continuum.  Irradiation/reprocessing can readily explain that correlation as the entire H$\alpha$ line is seen to respond quickly to X-ray variability.  Here too, we see that the X-ray and UV flux seem to respond simultaneously during the \swi\ observations (see Fig.~\ref{fig:swiuvxray}), though we do not have the data to be able to establish any lag.  This prompt response also supports irradiation/reprocessing as the origin of the correlation observed here.  However, we do not observe any clear correlation between the optical continuum and the X-ray flux (unlike in V404~Cyg), which would suggest that the UV and optical emitting regions are well-separated.  Alternatively, it could just be the case that in \cen, the optical flux is dominated by the donor star, and so any change in optical flux due to reprocessing may be too small to see. Furthermore, the longer orbital period of V404~Cyg \citep[6.5 days compared to 15 hours for Cen X-4;][]{casares92,cowley88} will lead to a larger disk, which would be more luminous at optical wavelengths.

The optical emission could be dominated by the donor star, as well as some contribution from the cooler outer regions of the truncated accretion disk.  The UV emission is too hot to arise from the donor star, and the X-ray/UV correlation suggests it must arise from a region that can see the central X-ray source, such as the inner region of the truncated accretion disk.  The correlation between X-ray and UV flux would seem to rule-out the accretion stream-impact point as the location for the UV emission, which had been previously suggested by \citet{mcclintock00} based on the \ion{Mg}{ii} $\lambda$2800 emission line in an {\it HST}/STIS spectrum of Cen~X-4 not showing a double-peak as would be expected for emission from the disk.  \citet{hynes12} closely examine the UV SEDs of 3 quiescent black hole and 1 quiescent neutron star and also rule out the stream-impact point as the origin of the UV emission.  They find that the mass accretion rate required to give the observed UV emission at the stream-impact point in the black hole GU~Mus and neutron star Aql~X-1 is 10 times higher than realistic average rates from these source based on their accretion histories.  They therefore favour an origin for the UV emission as being located in the hot inner region of the disk as also suggested by \citet{campanastella00} and \citet{mcclintock03}.  Our results also support such a picture.

It is important to consider whether there is a viable alternative to X-ray reprocessing as the origin of the X-ray/UV correlation.  For instance, could the emission in Cen~X-4 be jet-dominated and hence the UV flux and X-ray power-law both originate from the jet?  While radio emission has never been detected from a quiescent neutron star, it has been detected in quiescence in black hole X-ray binaries.  For instance, a radio detection of the black hole hole A0620$-$00 in quiescence \citep{gallo06} implies that there is a radio-emitting outflow during quiescence.  Broadband modeling of the quiescent SED in A0620$-$00 shows that the emission can be fitted with a maximally jet-dominated model with the jet emission dominating from radio through to the soft X-rays \citep{gallo07} and implies there are strong outflows \citep{froning11}.  Similarly, V404~Cyg has been detected at radio wavelengths in quiescence \citep{hjellming00,gallo05, hynes09}, with the radio spectrum consistent with synchrotron emission from a steady jet \citep{gallo05}.  In fact, it has been shown that quiescent black hole X-ray binaries could all be in jet-dominated states \citep{fender03}.  Neutron stars X-ray binaries, on the other hand, are known to be a factor of 30 or so fainter radio sources than black hole binaries during outburst \citep{fender01,migliari03}, and thus, if they follow the same dependence between the X-ray luminosity, $L_X$, and radio luminosity, $L_R$, then neutron stars could also be jet-dominated although at lower Eddington fractions \citep{fender03}.  However, neutron stars have been shown to have a different dependence between $L_X$ and $L_R$, implying that they never reach a jet-dominated state and remain X-ray dominated \citep{migliari06}. This therefore suggests that jet emission is unlikely to account for the X-ray/UV correlation observed here in Cen~X-4.

In conclusion, we have observed a significant X-ray/UV correlation in Cen~X-4, whose most likely explanation is due to irradiation of the inner edge of a truncated accretion disk by a central X-ray source leading to reprocessed UV emission.  Further X-ray/UV monitoring will help more firmly establish the link between X-ray and UV emission in Cen~X-4.

\section*{Acknowledgements} 
EMC thanks Rob Hynes, Edward Robinson and Federico Bernardini for helpful discussions on UV emission during quiescence.  RW acknowledges support from a European Research Council (ERC) starting grant.  N.D. is supported by NASA through Hubble postdoctoral fellowship grant number HSTHF-51287.01-A from the Space Telescope Science Institute. We acknowledge the use of public data from the Swift data archive.

\bibliographystyle{mn2e}
\bibliography{qNS}

\end{document}